\documentclass[12pt]{article}  
\usepackage{epsfig}
\usepackage{xspace}
\usepackage{cite}
\newcounter{enumct}
\newenvironment{Enumerate}{\begin{list}{\arabic{enumct}.}%
{\usecounter{enumct}\setlength{\topsep}{0.2mm}%
\setlength{\partopsep}{0.2mm}\setlength{\itemsep}{0.2mm}%
\setlength{\parsep}{0.2mm}}}{\end{list}}
\newcommand{\Y      }[3]{\ensuremath{#1\,^{+\,#2}_{-\,#3}}\xspace}
\newcommand{\epem   }{\ensuremath{\mathrm{e}^+\mathrm{e}^-}\xspace}
\newcommand{\aem    }{\ensuremath{\alpha}\xspace}
\newcommand{\ccbar  }{\ensuremath{\mathrm{c\bar{c}}}\xspace}
\newcommand{\qqbar  }{\ensuremath{\mathrm{q\bar{q}}}\xspace}
\newcommand{\invpb  }{\ensuremath{\mathrm{pb}^{-1}}\xspace}
\newcommand{\qsq    }{\ensuremath{Q^{2}}\xspace}
\newcommand{\wsq    }{\ensuremath{W^{2}}\xspace}

\newcommand{\psq    }{\ensuremath{P^{2}}\xspace}
\newcommand{\ft     }{\ensuremath{F_{2}^{\gamma}}\xspace}
\newcommand{\ftc    }{\ensuremath{F_{2,\mathrm{c}}^{\gamma}}\xspace}
\newcommand{\ftcxq  }{\ensuremath{\ftc(x,\qsq)}\xspace}
\newcommand{\ftcxqn }{\ensuremath{\ftcxq/\aem}\xspace}
\newcommand{\pt     }{\ensuremath{p_{\mathrm{t}}}\xspace}
\newcommand{\nch    }{\ensuremath{N_{\mathrm{ch}}}\xspace}
\newcommand{\dz     }{\ensuremath{{\mathrm{D}^{0}}}\xspace}
\newcommand{\ds     }{\ensuremath{{\mathrm{D}^{\star}}}\xspace}
\newcommand{\ttag   }{\ensuremath{\theta_{\rm tag}}\xspace}
\newcommand{\mev    }{\ensuremath{\mathrm{MeV}}\xspace}
\newcommand{\gev    }{\ensuremath{\mathrm{GeV}}\xspace}
\newcommand{\gevsq  }{\ensuremath{\mathrm{GeV}^2}\xspace}
\newcommand{\mc     }{\ensuremath{m_{\mathrm{c}}}\xspace}
\newcommand{\ssee   }{\ensuremath{\sqrt{s_{\rm e e}}}\xspace}
\newcommand{\eb     }{\ensuremath{E_\mathrm{b}}\xspace}
\newcommand{\ea     }{\ensuremath{E_\mathrm{a}}\xspace}
\newcommand{\etag   }{\ensuremath{E_\mathrm{tag}}\xspace}
\newcommand{\Wvis   }{\ensuremath{W_{\mathrm{vis}}}\xspace}
\newcommand{\xvis   }{\ensuremath{x_{\mathrm{vis}}}\xspace}
\newcommand{\etads  }{\ensuremath{\eta^{\ds}}\xspace}
\newcommand{\etadsa }{\ensuremath{\vert\etads\vert}\xspace}
\newcommand{\ptds   }{\ensuremath{\pt^{\ds}}\xspace}
\newcommand{\sigdsr }{\ensuremath{\sigma^{\ds}}\xspace}
\newcommand{\fcds   }{\ensuremath{f({\mathrm c}\to\ds)}\xspace}
\newcommand{\sigcc  }{\ensuremath{\sigma(\epem\to\epem\,\ccbar\,X)}\xspace}
\newcommand{\sigds  }{\ensuremath{\sigma(\epem\to\epem\,\ds\,X)}\xspace}
\newcommand{\Nrecds }{\ensuremath{N_{\rm rec}^{\ds}}\xspace}
\newcommand{\Ncords }{\ensuremath{N_{\rm cor}^{\ds}}\xspace}
\newcommand{\dm     }{\ensuremath{\Delta m}\xspace}
\newcommand{\dmn    }{\ensuremath{\Delta m_0}\xspace}
\newcommand{\RMC    }{\ensuremath{{\cal R}^{\mathrm{MC}}}\xspace}
\newcommand{\Nall   }{\ensuremath{55.3 \pm 11.0}\xspace}
\newcommand{\Nlow   }{\ensuremath{23.6 \pm  7.4}\xspace}
\newcommand{\Nhig   }{\ensuremath{31.4 \pm  8.1}\xspace}
\newcommand{\Slow   }{\ensuremath{ 3.1 \pm 1.0 \pm 0.5}\xspace}
\newcommand{\Shig   }{\ensuremath{ 2.6 \pm 0.9 \pm 0.3}\xspace}
\newcommand{\Sflow  }{\ensuremath{43.8 \pm 14.3 \pm 6.3 \pm 2.8}\xspace}
\newcommand{\Sfhig  }{\ensuremath{26.2 \pm  8.8 \pm 3.2 \pm 1.3}\xspace}
\newcommand{\SDhlow }{\ensuremath{34.5 \pm 14.3 \pm 6.9}\xspace}
\newcommand{\SNlow  }{\ensuremath{\Y{17.0}{2.9}{2.3}}\xspace}
\newcommand{\SNhlow }{\ensuremath{\Y{ 7.7}{2.2}{1.6}}\xspace}
\newcommand{\SNhig  }{\ensuremath{\Y{30.8}{7.1}{5.7}}\xspace}
\newcommand{\ftflow }{\ensuremath{0.180 \pm 0.059 \pm 0.026 \pm 0.012}\xspace}
\newcommand{\ftDhlow}{\ensuremath{0.154 \pm 0.059 \pm 0.029}\xspace}
\newcommand{\ftfhig }{\ensuremath{0.084 \pm 0.028 \pm 0.010 \pm 0.004}\xspace}
\newcommand{\ftNhlow}{\ensuremath{\Y{0.026}{0.007}{0.005}}\xspace}
\newcommand{\chiq   }{\ensuremath{\chi^2}\xspace}
\newcommand{\msqr   }{\ensuremath{\mu^{2}_{\rm r}}\xspace}
\newcommand{\msqf   }{\ensuremath{\mu^{2}_{\rm f}}\xspace}
\newcommand{\fcdsval}{\ensuremath{0.235 \pm 0.007 \pm 0.007}\xspace}
\newcommand{\BR     }{\ensuremath{7.66 \pm 0.22\,\%}\xspace}
\newcommand{\pb     }{\ensuremath{\mathrm{pb}}\xspace}
\begin{document}
\begin{titlepage}
\begin{center}{\large   EUROPEAN ORGANIZATION FOR NUCLEAR RESEARCH
}\end{center}\bigskip
\begin{flushright}
       CERN-EP-2002-031   \\ 10 May 2002 
\end{flushright}
\bigskip\bigskip\bigskip\bigskip\bigskip
\begin{center}{\huge\bf\boldmath
 Measurement of the Charm \\ Structure Function \ftc \\ of the Photon at LEP
 \unboldmath
}\end{center}\bigskip\bigskip
\begin{center}{\LARGE The OPAL Collaboration
}\end{center}\bigskip\bigskip
\bigskip\begin{center}{\large  Abstract}\end{center}
 The production of charm quarks is studied in deep-inelastic electron-photon
 scattering using data recorded by the OPAL detector at LEP at
 nominal \epem centre-of-mass energies from 183 to 209~GeV. 
 The charm quarks have been identified by full reconstruction of charged
 \ds mesons using their decays into $\dz\pi$ with the \dz observed in
 two decay modes with charged particle final states, 
 ${\rm K}\pi$ and ${\rm K}\pi\pi\pi$. 
 The cross-section \sigdsr for production of charged \ds 
 in the reaction $\epem\to\epem\,\ds\,X$ is measured in 
 a restricted kinematical region using two bins in Bjorken $x$, 
 $0.0014<x<0.1$ and $0.1<x<0.87$.
 From \sigdsr the charm production cross-section \sigcc and the charm 
 structure function of the photon \ftc are determined in the region 
 $0.0014<x<0.87$ and $5<\qsq <100$~\gevsq.
 For $x>0.1$ the perturbative QCD calculation at next-to-leading order
 agrees perfectly with the measured cross-section.
 For $x<0.1$ the measured cross-section is \Sflow~\pb
 with a next-to-leading order prediction of \SNlow~\pb.
\bigskip\bigskip\bigskip\bigskip
\bigskip\bigskip
\begin{center}{\large
(Submitted to Physics Letters B)
}\end{center}
\end{titlepage}
\begin{center}{\Large        The OPAL Collaboration
}\end{center}\bigskip
\begin{center}{
G.\thinspace Abbiendi$^{  2}$,
C.\thinspace Ainsley$^{  5}$,
P.F.\thinspace {\AA}kesson$^{  3}$,
G.\thinspace Alexander$^{ 22}$,
J.\thinspace Allison$^{ 16}$,
P.\thinspace Amaral$^{  9}$, 
G.\thinspace Anagnostou$^{  1}$,
K.J.\thinspace Anderson$^{  9}$,
S.\thinspace Arcelli$^{  2}$,
S.\thinspace Asai$^{ 23}$
% ...
}\end{center}\bigskip
\bigskip
$^{  1}$School of Physics and Astronomy, University of Birmingham,
Birmingham B15 2TT, UK
\newline
$^{  2}$Dipartimento di Fisica dell' Universit\`a di Bologna and INFN,
I-40126 Bologna, Italy
\newline
$^{  3}$Physikalisches Institut, Universit\"at Bonn,
D-53115 Bonn, Germany
\newline
% ...
$^{  5}$Cavendish Laboratory, Cambridge CB3 0HE, UK
\newline
% ...
$^{  9}$Enrico Fermi Institute and Department of Physics,
University of Chicago, Chicago IL 60637, USA
\newline
% ...
$^{ 16}$Department of Physics, Schuster Laboratory, The University,
Manchester M13 9PL, UK
\newline
% ...
$^{ 22}$Department of Physics and Astronomy, Tel Aviv University,
Tel Aviv 69978, Israel
\newline
$^{ 23}$International Centre for Elementary Particle Physics and
Department of Physics, University of Tokyo, Tokyo 113-0033, and
Kobe University, Kobe 657-8501, Japan
\newline
% ...
%
%
\section{Introduction}
\label{sec:intro}
 Much of the present knowledge of the structure of the photon has been
 obtained from deep-inelastic
 electron-photon\footnote{If not mentioned explicitly charge conjugation
 is implied, and for conciseness positrons are also referred 
 to as electrons. Natural units $\hbar=c=1$ are used throughout.} 
 scattering at \epem colliders~\cite{NIS-9904}.
 With the high statistics available at LEP2, it is possible to investigate 
 the flavour composition of the hadronic structure function \ft.
 The easiest flavour component of \ft to measure directly is the
 charm part \ftc, because charmed hadrons like the \ds meson are produced 
 with a large cross-section and can be identified by the well established
 method of full reconstruction of decays of charged \ds mesons.
 \par
 The measurement is based on deep-inelastic electron-photon scattering,
 ${\rm e}(k)\,\gamma(p) \rightarrow {\rm e}(k')\,\ccbar\,X$, 
 proceeding via the exchange of a virtual photon, $\gamma^*(q)$,
 (the symbols in brackets denote the four-momentum vectors of the particles).
 Experimentally, for these events one electron is observed in the detector
 together with a hadronic final state, and the second electron, which is only 
 slightly deflected, leaves the detector unobserved.
 The determination of \ftc exploits the fact that the differential 
 cross-section of this reaction as function of $\qsq=-q^2$ and Bjorken $x$,
 defined as $x={\qsq}/{(2p \cdot q)}$, is proportional to \ftcxq.
 \par
%
\begin{figure}[ht]
\begin{center}
{\includegraphics[width=0.6\linewidth]{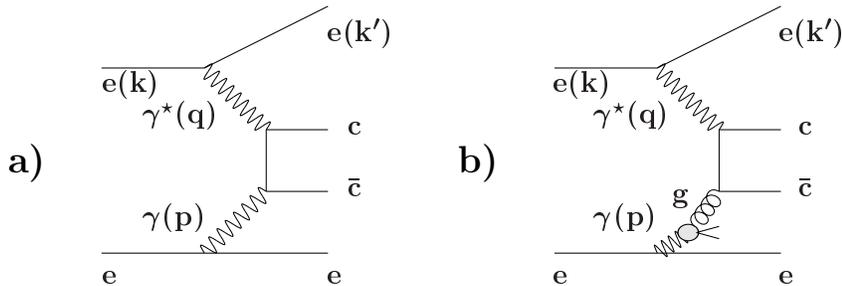}}
\caption{
         Examples of leading order diagrams contributing to
         a) the point-like, and b) the hadron-like part of \ftc.
        }\label{fig:fig01}
\begin{picture}(0,0)(0,0)
 \put(-170,100){\large {\bf a)}}\put(0,100){\large {\bf b)}}
\end{picture}
\end{center}
\end{figure}
%
 For a given gluon distribution function, the contribution to \ft from 
 charm quarks can be calculated in perturbative QCD, thanks to the 
 sufficiently large scale established by the charm mass, and predictions 
 have been evaluated at next-to-leading order (NLO) accuracy 
 in~\cite{LAE-9401LAE-9602}.
 As for light quarks, \ftc receives 
 contributions from the point-like and the hadron-like components of the 
 photon structure, as shown schematically in Figure~\ref{fig:fig01}. 
 These two contributions are predicted to have different dependences on
 $x$, with the hadron-like component dominating at very low values of $x$ and 
 the point-like part accounting for most of \ftc at $x>0.1$.
 \par
 The analysis presented here is an extension of the measurement of \ftc
 presented in~\cite{OPALPR294}, using basically the same analysis strategy,
 but a much larger data sample and refined Monte Carlo models.
 It is based on data recorded by the OPAL experiment in the years 1997--2000,
 with an integrated luminosity of ${\cal L}=654.1$~\invpb for nominal \epem 
 centre-of-mass energies, \ssee, from 183 to 209~GeV, with a luminosity 
 weighted average of $\ssee=196.6$~GeV.
%
%
\section{The OPAL detector}
\label{detec}
 A detailed description of the OPAL detector can be found in~\cite{OPALPR021},
 and therefore only a brief account of the main features relevant
 to the present analysis is given here.
 \par
 The central tracking system is located inside a solenoidal magnet which
 provides a uniform axial magnetic field of 0.435~T along the beam
 axis\footnote{In the OPAL coordinate system the $x$ axis points
 towards the centre of the LEP ring, the $y$ axis points upwards and
 the $z$ axis points in the direction of the electron beam.  
 The polar angle $\theta$ and the pseudorapidity $\eta=-\ln(\tan(\theta/2))$
 are defined with respect to the $z$ axis.}.
 The magnet is surrounded by a lead-glass electromagnetic
 calorimeter (ECAL) and a hadronic sampling calorimeter (HCAL).
 Outside the HCAL, the detector is surrounded by muon
 chambers. There are similar layers of detectors in the
 endcaps. The region around the beam pipe on both sides
 of the detector is covered by the forward calorimeters and the
 silicon-tungsten luminometers.
 \par
 Starting with the innermost components, the tracking system consists of 
 a high precision silicon microvertex detector~\cite{ALL-9401ALL-9301}, 
 a precision vertex drift chamber, a large volume jet chamber with 159 layers
 of axial anode wires, and a set of $z$ chambers measuring the track 
 coordinates along the beam direction.
 The transverse momenta \pt of tracks with respect to the $z$ direction of 
 the detector are measured with a precision of
 $\sigma_{\pt}/\pt=\sqrt{0.02^2+(0.0015\cdot \pt)^2}$ (\pt in GeV)
 in the central region, where $|\cos\theta|<0.73$. 
 The jet chamber also provides energy loss, ${\mathrm d}E/{\mathrm d}x$,
 measurements which are used for particle identification.
 \par
 The ECAL covers the complete azimuthal range for polar angles
 satisfying $|\cos\theta|<0.98$. The barrel section, which covers
 the polar angle range $|\cos\theta|<0.82$, consists of a cylindrical array
 of 9440 lead-glass blocks with a depth of
 $24.6$ radiation lengths. The endcap sections (EE) consist of 1132 lead-glass 
 blocks at each end with a depth of more than $22$ radiation lengths, 
 covering the polar angle region of $0.82 < |\cos\theta| < 0.98 $.
 \par
 The forward calorimeters (FD) at each end of the OPAL detector
 consist of cylindrical lead-scintillator calorimeters with a depth of
 24 radiation lengths divided azimuthally into 16 segments.
 The electromagnetic energy resolution is about
 $18\%/\sqrt{E}$~($E$ in GeV).
 The acceptance of the forward calorimeters covers the angular range
 between 47 and 140~mrad from the beam direction.
 Three planes of proportional tube chambers at 4 radiation lengths
 depth in the calorimeter measure the direction of electron showers
 with a precision of approximately 1~mrad.
 \par
 The silicon tungsten detectors (SW)~\cite{AND-9401} are located in front 
 of the forward calorimeters at each end of the OPAL detector.
 Their clear acceptance covers a polar angular region between 33 and 59~mrad.
 Each calorimeter consists of 19 layers of silicon detectors and 18
 layers of tungsten, corresponding to a total of 22 radiation
 lengths. Each silicon layer consists of 16 wedge-shaped 
 silicon detectors. The electromagnetic energy resolution is about
 $25\%/\sqrt{E}$ ($E$ in GeV). The radial position of electron
 showers in the SW calorimeter can be determined with a
 typical resolution of 0.06 mrad in the polar angle $\theta$.
 The SW detector provides the luminosity measurement. 
%
%
\section{Monte Carlo simulation}
\label{sec:MC}
 The Monte Carlo models HERWIG6.1~\cite{MAR-9201} and 
 PYTHIA6.1~\cite{SJO-0101}, both based on leading order (LO) matrix 
 elements and parton showers, are used to model the deep-inelastic
 electron-photon scattering events $\epem\to\epem\ccbar\,X$.
 For both Monte Carlo models, the charm quark mass is chosen to be 
 $\mc=1.5$~GeV.
 \par
 In HERWIG6.1, the cross-section is evaluated separately for the 
 point-like and hadron-like contributions using matrix elements for 
 massive quarks, together with the GRV parametrisation~\cite{GLU-9201GLU-9202}
 for the gluon distribution of the photon\footnote{%
 The massive matrix elements for the point-like contribution have been
 implemented into the standard HERWIG6.1 model by J.~Ch\'{y}la.
 The hadron-like component is based on the matrix elements of the boson-gluon 
 fusion process. 
 }.
 This is an improvement compared to the analysis presented in~\cite{OPALPR294},
 where HERWIG5.9 was used with matrix elements for massless quarks only.
 For that model the effect of the charm quark mass was accounted for
 only rather crudely by not simulating events with hadronic masses of
 less than $2m_{\rm c}$.
 The fragmentation into hadrons is based on the cluster model for 
 HERWIG6.1, using the OPAL tune\footnote{The main changes are that
 meson states that do not belong to the $L=0,1$ supermultiplets
 are removed, and that the parameters CLSMR(1), PSPLT(2) and DECWT 
 have been changed from their default values of 0.0, 1.0, and 1.0 to 
 0.4, 0.33, and 0.7. A detailed description of the tune can be 
 obtained from the HERWIG web interface.}.
 \par
 To obtain an estimate of the dependence of the measurements on
 the details of the modelling of \ds production and fragmentation,
 a second model, based on the PYTHIA program, is used. 
 The spectrum of photons with varying virtualities 
 is generated in PYTHIA~\cite{Friberg:1999zz}. 
 The point-like and the hadron-like processes,
 which in PYTHIA are denoted by direct and resolved,
 are then simulated separately using the matrix elements for the
 production of massive charm quarks, $\gamma\gamma\to \ccbar$
 (subprocess ISUB=85) and $\gamma {\rm g}\to\ccbar$ (subprocess ISUB=84).
 Since these matrix elements are valid only for real photons ($Q^2=0$) 
 the $Q^2$ dependence of the cross-sections is not expected
 to be correctly modelled. The cross-sections for both
 contributions are therefore taken from HERWIG, and PYTHIA is
 only used for shape comparisons. 
 In PYTHIA the fragmentation into hadrons is simulated using the Lund 
 string model.
 \par
 All Monte Carlo events were passed through the
 full simulation of the OPAL detector~\cite{ALL-9201}.
 They are analysed using the same reconstruction algorithms as
 applied to the data.
%
%
\section{Kinematics and data selection}
\label{sec:evsel}
 To measure \ftcxq, the distribution of events in $x$ and \qsq is needed. 
 These variables are related to the experimentally measurable quantities
 \etag and \ttag by:
%
\begin{equation}
  \qsq = 2\,\eb\,\etag\,(1- \cos\ttag)
  \label{eqn:qsq}
\end{equation}
%
 and
%
\begin{equation}
  x = \frac{\qsq}{\qsq+W^2+\psq}\,,
  \label{eqn:Xcalc}
\end{equation}
%
 where \eb is the energy of the beam electrons, \etag and \ttag are the 
 energy and polar angle of the deeply inelastically scattered
 electron, \wsq is the invariant mass squared of the hadronic final state,
 and $\psq=-p^2$ is the negative value of the virtuality squared of the
 quasi-real photon.
 It is required that the associated electron is not seen in the detector.
 This ensures that $\psq \ll \qsq+\wsq$, therefore \psq is neglected when 
 calculating $x$ from Equation~\ref{eqn:Xcalc}.
 The electron mass is neglected throughout. 
 \par
 Deep-inelastic electron-photon scattering candidate events are selected as
 follows:
%
\begin{Enumerate} 
 \item The calorimeter cluster with the highest energy in either SW 
       or FD is taken as the electron candidate. The polar angle
       \ttag is measured with respect to the original beam direction.
       An electron candidate is required to have $\etag>0.5\eb$ and 
       $33 <\ttag < 55$~mrad (SW) or $60 <\ttag < 120$~mrad (FD).
 \item To ensure that the virtuality of the quasi-real photon is small, 
       the highest energy electromagnetic cluster in the SW and FD detectors
       in the hemisphere opposite the scattered electron must have an 
       energy $\ea \le 0.25\,\eb$ (the anti-tag condition).
 \item At least three tracks must be found in the tracking chambers.
       A track is required to have a minimum transverse momentum 
       with respect to the beam axis
       of 0.12~\gev, and to fulfill standard quality criteria~\cite{OPALPR294}.
 \item To reduce background from \epem annihilation events with charged \ds 
       mesons in the final state, the sum of the energy of all calorimeter
       clusters in the ECAL is required to be less than 0.5\eb. Electromagnetic
       calorimeter clusters have to pass an energy threshold of 0.1~\gev
       for the barrel section and 0.25~\gev for the endcap sections.
 \item To reduce the \epem annihilation background further, the visible 
       invariant mass of the event, \Wvis, should be less than 0.65\eb. 
       The invariant mass \Wvis is calculated using the momenta of tracks 
       and using the energies and positions of clusters measured in the ECAL, 
       the HCAL, the FD and the SW calorimeters. Clusters in the 
       SW and FD in the hemisphere of the tagged electron are excluded. 
       A matching algorithm~\cite{OPALPR150} is applied to avoid double 
       counting the energy of particles that produce both tracks and clusters.
       All tracks, except for the kaon candidate identified in the \ds
       reconstruction, are assumed to be pions.
 \item A fully reconstructed charged \ds candidate has to be present.
       The transverse momentum of the \ds with respect to the beam axis, 
       \ptds, should fulfill $\ptds>1(3)$~\gev for SW(FD)-tagged events, and
       the pseudorapidity \etads should be in the range $\etadsa<1.5$.
       The \ds mesons are identified using their decay into $\dz\pi$ 
       with the \dz observed in two decay modes with charged particle
       final states, ${\rm K}\pi$ and ${\rm K}\pi\pi\pi$. 
       The following quality criteria are applied to the \dz candidate.
       For the kaon candidate, the ${\mathrm d}E/{\mathrm d}x$ probability 
       for the kaon hypothesis
       should
       exceed 10$\%$, and for all pion candidates the pion probability
       should be above 0.5$\%$.
       In addition, the mass of the \dz candidate should lie in the window
       1.79--1.94~\gev, and the cosine of the angle between the kaon 
       momentum in the \dz rest frame and the \dz momentum in the 
       laboratory frame should be below 0.9.
       The cut on the cosine of the angle reduces the background
       which peaks at 1, whereas the pseudo-scalar \dz decays isotropically 
       in its rest frame.
\end{Enumerate}
%
 With these cuts 1653 events are selected in the data having $\dm<0.21$~\gev,
 where \dm is the difference between the \ds and \dz candidate masses.
 They are shown in Figure~\ref{fig:fig02}.
 The number of events reconstructed in the signal region defined by
 $0.1424<\dm<0.1484$~\gev is 115.
 No event with more than one \ds candidate in the signal region has been
 found in the data.
 \par
 Using Monte Carlo simulations, the expected background to the 
 deep-inelastic electron-photon scattering events from all other Standard 
 Model physics processes that potentially contain final state \ds mesons is
 found to be about one event and is neglected.
 The background from random coincidences between
 off-momentum\footnote{Off-momentum electrons originate from beam gas
 interactions  far from the OPAL interaction region and are 
 deflected into the detector by the focusing quadrupoles.} 
 beam electrons faking a scattered electron and 
 photon-photon scattering events without an observed electron
 has been estimated to be below 1.4 events and has been neglected.
 Thus, only the combinatorial background from deep-inelastic electron-photon 
 scattering events $\epem\to\epem\qqbar$ with ${\rm q} = {\rm u}, {\rm d},
 {\rm s} \mbox { and } {\rm c}$ has to be subtracted from the data.
 The production of bottom quarks is suppressed compared to charm
 production due to the larger mass and smaller charge and is 
 negligible~\cite{NIS-9904}.
 \par
%
%
\section{Results}
\label{sec:resu}
%
%
\subsection{Comparison of data and theoretical predictions}
\label{sec:comp}
 Figure~\ref{fig:fig02} shows the difference between the \ds and 
 \dz candidate masses for both decay channels combined.
 A clear peak is observed around the mass difference between the \ds and 
 the \dz meson, which is $\dmn = 0.1454$~\gev~\cite{PDG-0001}.
 The number of signal events, \Nrecds, has been obtained from an unbinned
 maximum likelihood fit to this distribution.
 Similar fits are performed in two regions of \xvis, calculated from 
 Equation~\ref{eqn:Xcalc} using \Wvis and the measured value of \qsq.
 The fit function contains a Gaussian for the signal and a power law function
 of the form $a\cdot(\dm-m_{\pi})^b$, for the background contribution.
 Because the number of signal events is small, the mean value of the Gaussian 
 is kept fixed to the PDG value in the fit.
 The width and the normalisation are determined by the fit, which 
 gives \Nall signal events, with \Nlow and \Nhig events for $\xvis<0.1$ 
 and $\xvis>0.1$, where the uncertainties are statistical.
 The fitted width is $0.70\pm 0.16$~\mev when using the whole sample.
 The quality of the fit is satisfactory.
 The \chiq between the fitted curve and the data in Figure~\ref{fig:fig02}
 is 94 for 71 non-empty bins.
 The corresponding \chiq for events with $\xvis<0.1$ and $\xvis>0.1$
 are 71 and 93.
 \par
 The result of the fit for all events is shown in Figure~\ref{fig:fig02}
 together with the absolute prediction of the combinatorial
 background measured from the data using events with a wrong-charge 
 pion for the $\ds\to\dz\pi$ decays.
 The fit agrees with this second estimate of the combinatorial background.
 The numbers of signal events for $\xvis<0.1$ and $\xvis>0.1$ predicted by
 the HERWIG model are $12.1 \pm 0.6$ and $30.0 \pm 1.1$, 
 where the uncertainties are statistical.
 \par
 Figure~\ref{fig:fig03} shows, for the signal events, the distributions of 
 four global event quantities, \qsq, \Wvis, \xvis and the charged multiplicity
 \nch, and two variables related to the kinematics of the \ds candidates, 
 \ptds and \etadsa.
 The signal events are shown for the \dm region $0.1424<\dm<0.1484$~\gev, 
 after subtracting the combinatorial background in that region.
 The normalisation of the background is given by the result of the fit
 for the complete sample.
 The shape of the background distributions is taken from the data using 
 both the events with the wrong-charge combinations and the events
 with the correct charge combinations that fulfill $0.16<\dm<0.21$~\gev, 
 see Figure~\ref{fig:fig02}.
 It has been verified from Monte Carlo that in this range the distributions 
 shown are independent of \dm.
 Subtracting the background this way is statistically more precise than using
 only the events with the wrong-charge combinations in the signal region.
 The data are compared to the prediction of the HERWIG and PYTHIA
 Monte Carlo models, normalised to the data luminosity.
 For the HERWIG prediction the hadron-like component (HERWIG HL) is shown 
 on top of the point-like component (HERWIG PL).
 Overall, the HERWIG model describes the data distributions quite well, and
 the shape of the PYTHIA prediction is also consistent with the data.
 \par
 In what follows, the cross-sections has been calculated based on the HERWIG
 prediction, while the PYTHIA program was used as a second Monte Carlo model
 to estimate the uncertainty stemming from the Monte Carlo description of the 
 data.
%
%
\subsection{Determination of the charm production cross-section}
\label{sec:cross}
 The inclusive cross-section \sigdsr has been extracted in a well-measured 
 kinematic region where $\ptds>1$~\gev for an electron scattering angle of 
 $33-55$~mrad, or $\ptds>3$~\gev for $60-120$~mrad,
 $\etadsa<1.5$ and $5<\qsq<100$~\gevsq, 
 using almost the whole accessible \qsq range defined by \ttag and \etag.
 \par
 As in the previous investigation~\cite{OPALPR294}, the analysis is 
 performed in two bins in $x$, $0.0014<x<0.1$ and $0.1<x<0.87$.
 The $x$ range is limited by the \qsq range, by the minimum kinematically
 allowed invariant mass $W=3.88$~\gev needed for the production of a \ds 
 meson together with a second charmed hadron, and by the event selection 
 cut on \Wvis.
 To take into account the detector acceptance and resolution in $x$
 the data are corrected using a $2\times2$ matrix, which contains the 
 information of the correlation between the measured value \xvis and the 
 generated $x$ for the two bins in $x$ as given by Monte Carlo.
 The effects due to the migration in \qsq are small compared to those
 in $x$, and are neglected.
 \par
 For a given \ds decay channel the selection efficiency for $x<0.1$ is given 
 by the ratio of the number of reconstructed \ds mesons originating 
 from events with $x<0.1$ to all generated \ds mesons in that channel
 with $x<0.1$ in the restricted kinematic region defined above.
 For HERWIG6.1, the selection efficiencies for $x<0.1$ are $(16 \pm 1)\%$
 and $(12 \pm 1)\%$ for the point-like and hadron-like components. 
 For $x>0.1$ they are $(19 \pm 1)\%$ and $(19 \pm 5)\%$ respectively.
 Combining the two components yields efficiencies of $(15 \pm 1)\%$ and 
 $(19 \pm 1)\%$ for $x<0.1$ and $x>0.1$.
 The corresponding selection efficiencies for the PYTHIA sample are 
 $(14 \pm 1)\%$ and $(20 \pm 1)\%$.
 This means that in each bin the predicted efficiencies from the HERWIG and
 PYTHIA models are consistent, so the size of their error of about 
 10$\%$ is taken as the systematic uncertainty on the efficiency.
 \par
 Table~\ref{tab:result}a) summarises the measured values of \sigdsr
 for \ds production in the restricted kinematic region defined above:
%
 \begin{equation} 
 \sigdsr=\frac{\Ncords}{{\rm BR} \cdot {\cal L}}.
 \end{equation}
%
 It is calculated from the number of \ds events, \Ncords, obtained from
 \Nrecds determined by the likelihood fit to the full sample, together with
 the $2\times 2$ matrix correction based on the HERWIG model.
 The statistical correlation between the two data points is $-28\%$.
 For the combined \ds branching ratios for the 3-prong and 
 5-prong decay modes, ${\rm BR}=\BR$~\cite{PDG-0001} is used.
 \par
 The present measurement makes use of the estimation of the systematic error
 in~\cite{OPALPR294} with some small changes. 
 The statistical error from the background subtraction is treated differently
 and is now included in the statistical error of the number of signal events 
 determined by the fit.
 For each of the $x$ regions, a total correlated systematic uncertainty of 
 10$\%$, as evaluated in~\cite{OPALPR294}, is attributed to the sum of the 
 uncertainties stemming from the branching ratios, the imperfectness
 of the modelling of the central tracking detectors, the kaon identification 
 and the variation of the energy scale of the electron candidate.
 \par
 For the determination of the cross-section of charm production, \sigcc, the
 Monte Carlo models are used for extrapolation.
 This allows \sigcc to be calculated via the relation 
%
\begin{eqnarray}
\sigcc  & = & \frac{1}{2\cdot\fcds} \cdot \sigds \nonumber\\
        & = & \frac{1}{2\cdot\fcds} \cdot \RMC \cdot \sigdsr .
\label{eq-sigtag}
\end{eqnarray}
%
 The value used for the charm to \ds hadronisation fraction, \fcds, is
 taken to be $\fcdsval$, independent of $x$.
 It has been estimated in~\cite{GLA-9901} by averaging results
 obtained in \epem annihilation at LEP1 as well as at lower 
 \epem centre-of-mass energies.
 The extrapolation factor \RMC is defined as the ratio of the number of
 all generated \ds mesons in the full kinematic region of \ptds and \etadsa
 divided by the number of generated \ds mesons in the restricted region.
 For $x<0.1$ the \RMC values for the HERWIG and PYTHIA models are
 $6.15 \pm  0.03$ and $6.23 \pm  0.03$, while for $x>0.1$ the corresponding 
 numbers are $4.73 \pm  0.02$ and $4.59 \pm  0.02$, where the errors are 
 statistical. 
 The central values of the extrapolated cross-sections are obtained using
 the HERWIG numbers, and the differences between the HERWIG and PYTHIA
 predictions have been taken as estimates of the extrapolation errors.
 \par
 In Figure~\ref{fig:fig04}a) and Table~\ref{tab:result}b), the measured
 cross-sections \sigcc are compared to the calculation 
 of~\cite{LAE-9401LAE-9602} performed in LO and NLO and to the
 prediction from HERWIG.
 The NLO prediction uses $\mc=1.5$~\gev. The renormalisation 
 and factorisation scales are chosen to be $\msqr = \msqf = \qsq$. 
 The calculation is obtained for the sum of the point-like and
 hadron-like contributions to \ftc, where for the calculation of the 
 hadron-like part the GRV-NLO parametrisation is used.
 These GRV-NLO parton distributions of the photon have been found 
 to describe the OPAL jet data~\cite{OPALPR250}.
 The NLO corrections are predicted to be small for the whole $x$ range.
 The NLO calculation is shown in Figure~\ref{fig:fig04}
 as a band representing the uncertainty of 
 the theoretical prediction, evaluated by varying the charm quark mass
 between 1.3 and 1.7~GeV and by changing the renormalisation and
 factorisation scales in the range $\qsq/4 \le\msqr=\msqf\le 4\qsq$,
 taking the largest difference from the central value as the error.
 The HERWIG prediction falls within the uncertainty band of the NLO
 calculation.
 \par
 For $x>0.1$, the predicted NLO cross-section in Figure~\ref{fig:fig04}a)
 agrees well with the data.
 For $x<0.1$, the situation is different. The NLO calculation predicts
 the hadron-like and point-like component to be of about equal size, and
 the sum is smaller than what is observed for the data.
 \par
 The point-like contribution can be calculated with small uncertainties,
 i.e.~in LO it depends only on \mc. 
 The prediction for the hadron-like part is more uncertain, especially
 because it depends on the gluon distribution of the photon, for 
 which experimental information is limited.
 Given the good agreement at large $x$ between the data and the NLO
 prediction, dominated by the point-like component, the NLO point-like
 prediction has been subtracted from the measured cross-section in the
 region $0.0014<x<0.1$.
 This leads to the value of \SDhlow~\pb for the hadron-like contribution 
 to the cross-section, which is to be compared with the NLO prediction of 
 \SNhlow~\pb, when using the GRV-NLO parton distributions of the photon.
%
%
\subsection{Extraction of \boldmath\ftc\unboldmath}
\label{sec:f2ex}
 The value of the charm structure function \ftcxqn of the photon, averaged 
 over the corresponding bin in $x$, and given at a fixed value of \qsq,
 is determined by 
%
\begin{equation}
 \ftcxqn=\sigcc\cdot\left(\frac{\ftcxqn}{\sigcc}\right)_{\rm NLO},
 \label{eq-nlo}
\end{equation}
%
 where the ratio $(\ftcxqn/\sigcc)_{\rm NLO}$ is given
 by the NLO calculation of~\cite{LAE-9401LAE-9602}.
 This approach assumes that for a bin of $x$ the ratio of the
 structure function, averaged over $x$ and evaluated at the average
 \qsq value of the data, and the cross-section within the region of
 \qsq, are the same for the data and the NLO calculation.
 The measurement is given at $\qsq = 20$~\gevsq, which roughly corresponds
 to the average value of \qsq observed for the data.
 The \ftcxqn values are listed in Table~\ref{tab:result}c) and shown
 in Figure~\ref{fig:fig04}b) on a logarithmic scale in $x$.
 \par 
 In addition to the structure function of the full NLO calculation, the 
 predicted hadron-like component is also shown in Figure~\ref{fig:fig04}b).
 This contribution is very small for $x>0.1$ and therefore in this 
 region the NLO calculation is a perturbative prediction which depends only
 on the charm quark mass and the strong coupling constant.
 This prediction agrees perfectly with the data.
 \par
 To illustrate the shape of \ftc the data are also compared to the
 GRS-LO~\cite{GLU-9501} prediction and to the point-like component alone,
 both shown for $\qsq=20$~\gevsq.
 The point-like contribution is small at low $x$.
 The data span a rather large range in \qsq, in which the change of the 
 predicted \ftc is large.
 The maximum value of \ftc for $x>0.1$ rises by about a factor of five
 between the lower boundary of $\qsq=5$~\gevsq and the higher boundary of 
 $\qsq=100$~\gevsq.
 In addition, the mass threshold, $W=2\mc$, introduces a \qsq dependent
 upper limit in $x$ which, in the \qsq range studied, varies from
 $x=0.35$ to $x=0.9$.
 \par
 To be able to compare the data directly with the curves from 
 the GRS-LO predictions the points are placed at those $x$ positions 
 that correspond to the average predicted \ftc within the 
 bin~\cite{Lafferty}.
 These $x$ values are calculated both for the full \ftc and for the 
 point-like component alone.
 The data points are located at the mean of the two values and the
 horizontal error bar indicates their difference.
 For $x>0.1$ the difference is invisible in Figure~\ref{fig:fig04}b).
 \par
 For $x<0.1$ the hadron-like contribution to \ftcxqn, determined as for 
 the cross-section, amounts to \ftDhlow, to be compared with the NLO 
 prediction of \ftNhlow.
%
%
\section{Conclusions}
\label{sec:concl}
 The production of charm quarks is studied in deep-inelastic
 electron-photon scattering using data recorded by the OPAL detector
 at LEP at nominal \epem centre-of-mass energies from 183 to 209~GeV 
 in the years 1997--2000.
 The result is based on 654.1~\invpb of data with a luminosity 
 weighted average centre-of-mass energy of $196.6$~GeV.
 The measurement is an extension of the result presented in~\cite{OPALPR294},
 using basically the same analysis strategy, but with improved Monte Carlo
 models and higher statistics. 
 The two OPAL results are consistent and the new measurement supersedes
 the result in~\cite{OPALPR294}.
 \par
 The cross-section \sigdsr for \ds production in the 
 reaction $\epem\to\epem\,\ds\,X$ is measured in the deep-inelastic 
 scattering regime for a restricted region
 where $\ptds>1$~\gev for an electron scattering angle of 
 $33-55$~mrad, or $\ptds>3$~\gev for $60-120$~mrad,
 $\etadsa<1.5$ and $5<\qsq<100$~\gevsq, divided into two bins 
 in $x$, $0.0014<x<0.1$ and $0.1<x<0.87$.
 Within errors the cross-sections can be described by the HERWIG model.
 From \sigdsr the cross-section \sigcc is obtained by extrapolation in 
 the same bins of $x$ and \qsq, using Monte Carlo.
 \par
 The charm structure function \ftcxqn of the photon is evaluated 
 at $\qsq=20$~\gevsq for the same bins of $x$.
 For $x>0.1$, the perturbative QCD calculation at next-to-leading order 
 agrees perfectly with the measurement.
 For $x<0.1$ the point-like component, however calculated, lies below
 the data. Subtracting the NLO point-like prediction, a measured
 value for the hadron-like part of \ftDhlow is obtained.
%
%
\appendix
\par
\section*{Acknowledgements:}
\par
 We especially wish to thank Eric Laenen for many interesting and
 valuable discussions and for providing the software to calculate
 the NLO predictions. 
 We are grateful to J.~Ch\'{y}la for the implementation of the point-like
 contribution for massive matrix elements into the HERWIG framework.
 \par
 We particularly wish to thank the SL Division for the efficient operation
 of the LEP accelerator at all energies and for their close cooperation with
 our experimental group. In addition to the support staff at our own
 institutions we are pleased to acknowledge the  \\
 Department of Energy, USA, \\
 National Science Foundation, USA, \\
 Particle Physics and Astronomy Research Council, UK, \\
 Natural Sciences and Engineering Research Council, Canada, \\
 Israel Science Foundation, administered by the Israel
 Academy of Science and Humanities, \\
 Benoziyo Center for High Energy Physics,\\
 Japanese Ministry of Education, Culture, Sports, Science and
 Technology (MEXT) and a grant under the MEXT International
 Science Research Program,\\
 Japanese Society for the Promotion of Science (JSPS),\\
 German Israeli Bi-national Science Foundation (GIF), \\
 Bundesministerium f\"ur Bildung und Forschung, Germany, \\
 National Research Council of Canada, \\
 Hungarian Foundation for Scientific Research, OTKA T-029328, 
 and T-038240,\\
 Fund for Scientific Research, Flanders, F.W.O.-Vlaanderen, Belgium.\\
%
%

%
\clearpage
\renewcommand{\arraystretch}{1.20}
\begin{table}[t]
\small
\begin{center}
\begin{tabular}{|c|c|c|}
\cline{2-3}
\multicolumn{1}{c}{{\large a)}}&\multicolumn{2}{|c|}{\sigdsr [pb]}\\\cline{2-3}
\multicolumn{1}{c|}{}&$0.0014<x<0.1$&$0.1<x<0.87$\\\hline
OPAL          &              \Slow &              \Shig  \\
HERWIG        & 1.13 (0.71 + 0.43) &  2.05 (2.02 + 0.03) \\
\hline\multicolumn{3}{c}{}\\
\cline{2-3}
\multicolumn{1}{c}{{\large b)}}&\multicolumn{2}{|c|}{\sigcc [pb]}\\\cline{2-3}
\multicolumn{1}{c|}{}&$0.0014<x<0.1$&$0.1<x<0.87$\\\hline
OPAL          &             \Sflow &             \Sfhig \\
HERWIG        &  17.0 (8.5 +\phantom{3}8.4)  & 27.4 (27.1 + 0.4) \\
LO            &  16.0 (8.1 +\phantom{3}7.9)  & 26.7 (26.3 + 0.4) \\
NLO           & \SNlow (\Y{ 9.3}{0.8}{0.6} + \SNhlow)
              & \SNhig (\Y{30.4}{6.9}{5.5} + \Y{0.4}{0.2}{0.1})\\
\hline\multicolumn{3}{c}{}\\
\cline{2-3}
\multicolumn{1}{c}{{\large c)}}&\multicolumn{2}{|c|}{\ftcxqn}\\\cline{2-3}
\multicolumn{1}{c|}{}&$0.0014<x<0.1$&$0.1<x<0.87$\\\hline
OPAL          &                \ftflow & \ftfhig                 \\
LO            &                  0.065 &                   0.084 \\
NLO           &\Y{0.070}{0.011}{0.008} (\Y{0.044}{0.004}{0.003} +
                                        \ftNhlow ) 
              &\Y{0.099}{0.023}{0.019} (\Y{0.098}{0.023}{0.019} + 0.001) \\
\hline
\end{tabular}
\normalsize
\caption{a) The cross-section \sigdsr in the restricted region,
         b) the cross-section \sigcc, extrapolated using Monte Carlo,
         and
         c) the structure function \ftc at $\qsq=20$~\gevsq.
         For the measurements the quoted errors are due, respectively,
         to statistics, systematics and extrapolation.
         The statistical correlation between the two data points is $-28\%$. 
         The measurements are compared to several theoretical predictions.
         The HERWIG prediction in a) uses the value of \fcds contained in 
         the program.
         The numbers in brackets are separate values for the point-like and
         hadron-like components, respectively. See text for further details.
        }\label{tab:result}
\end{center}\end{table}
%
%
\begin{figure}[tbp]
\begin{center}
{\includegraphics[width=1.0\linewidth]{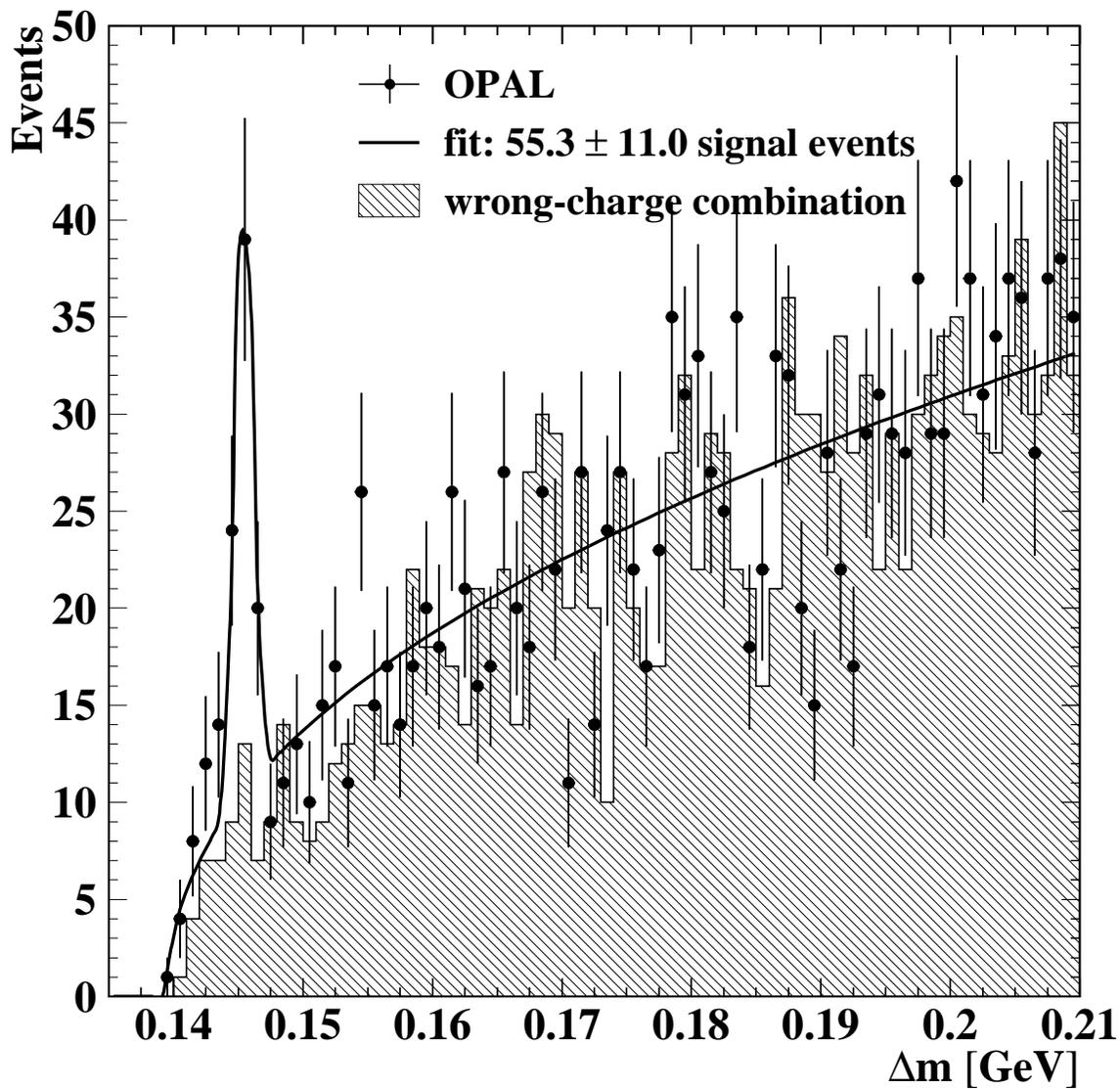}}
\caption{Distribution of the difference, \dm, between the  
         \ds and \dz candidate masses for both decay channels combined.
         The data are shown as points with statistical errors. 
         The histogram represents the combinatorial background estimated 
         using events with a wrong-charge pion for the $\ds\to\dz\pi$ decays.
         The curve is the result of the fit to a Gaussian signal plus a 
         power-law background function, as explained in the text.
        }\label{fig:fig02}
\end{center}
\end{figure}
%
\begin{figure}[tbp]
\begin{center}
{\includegraphics[width=1.1\linewidth]{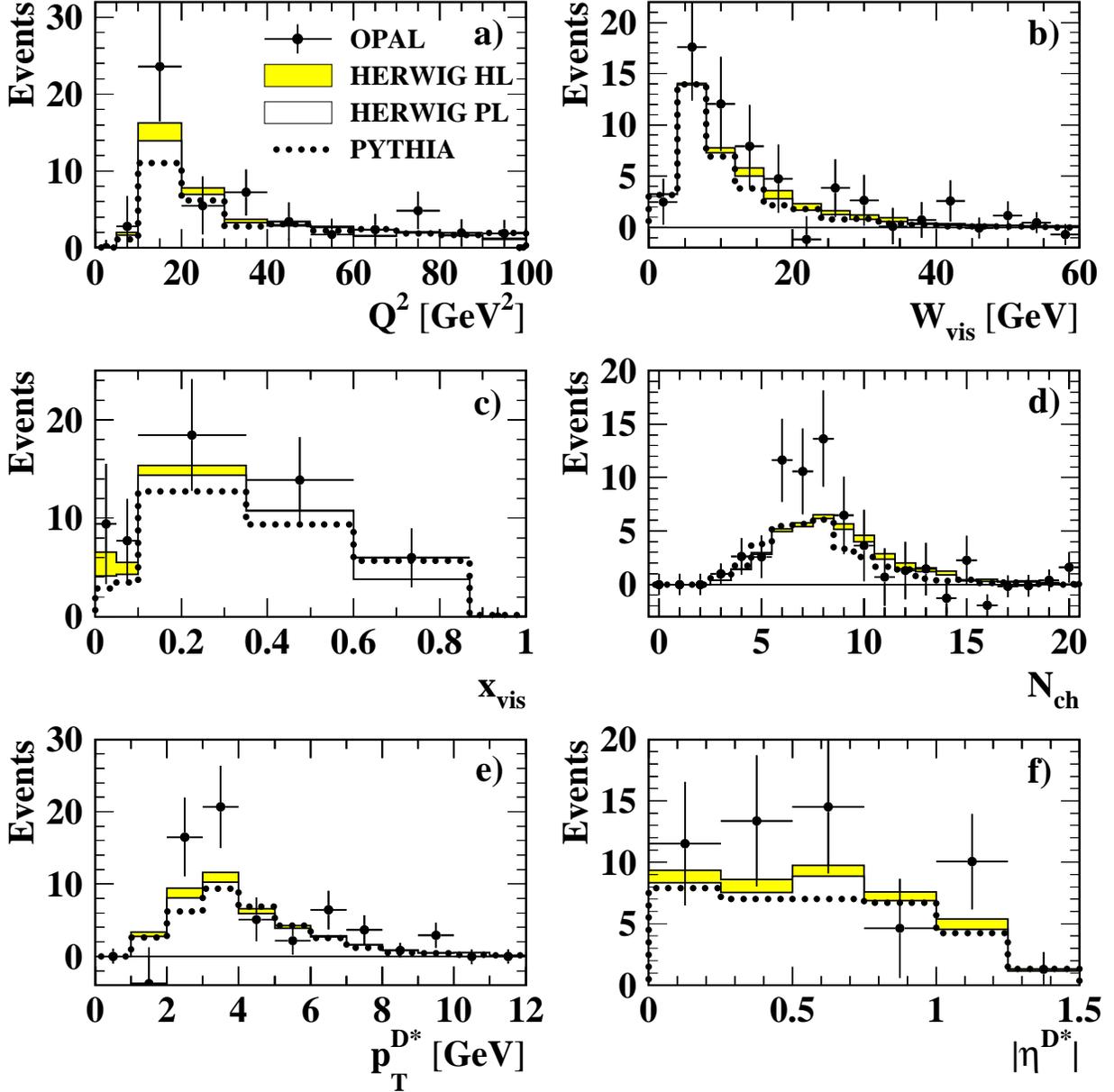}}
\caption{Background subtracted data distributions compared to predictions 
         based on the HERWIG and PYTHIA programs. 
         For HERWIG the hadron-like component is shown as a shaded histogram
         on top of the open histogram representing the point-like component.
         The Monte Carlo predictions are normalised to the data luminosity.
         Shown are 
         a) \qsq, 
         b) the visible invariant mass, \Wvis,
         c) \xvis,
         d) the charged multiplicity, \nch,
         e) the transverse momentum of the \ds with respect to the beam axis,
         \ptds, and 
         f) the absolute value of the pseudorapidity \etadsa.
         The data are shown by points with error bars representing the 
         statistical error.
        }\label{fig:fig03}
\end{center}
\end{figure}
%
\begin{figure}[tbp]
\begin{center}
{\includegraphics[width=0.9\linewidth]{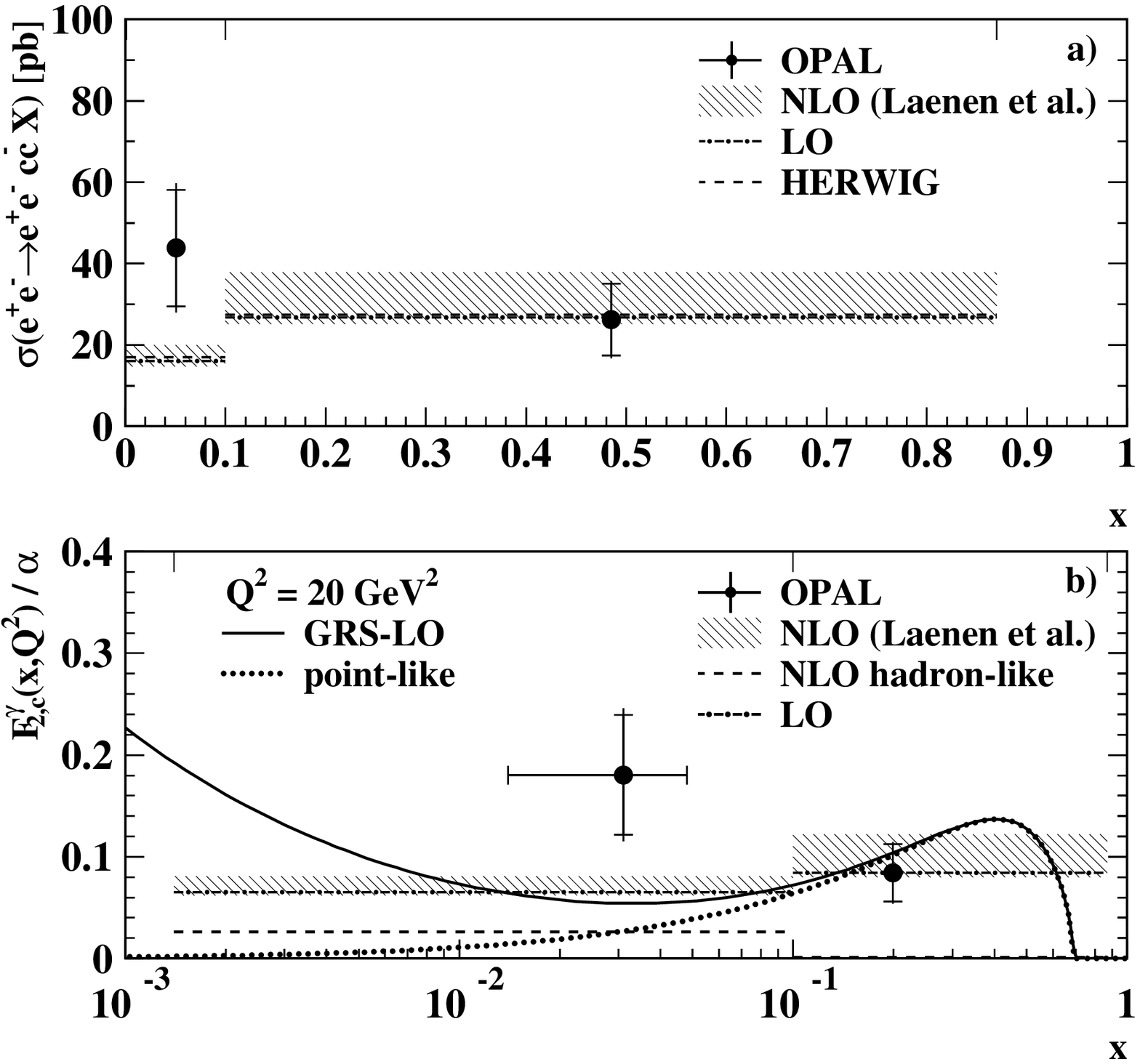}}
\caption{OPAL results for a) the cross-section \sigcc, with 
         $5<\qsq<100$~\gevsq and
         b) for the charm structure function of the photon divided
         by the fine structure constant, \ftcxqn, at $\qsq=20$~\gevsq.
         The data points are the results obtained
         with the HERWIG Monte Carlo model.
         The outer error bar is the total error and the inner error
         bar the statistical error.       
         The data points in b) are placed at those $x$ values that correspond
         to the average predicted \ftc within a bin.
         The data are compared to the calculation
         of~\protect\cite{LAE-9401LAE-9602} performed in LO and NLO. 
         The band for the NLO calculation
         indicates the theoretical error from uncertainties in the
         charm quark mass and renormalisation and factorisation scales. 
         In a) the cross-section prediction of the HERWIG Monte Carlo
         model is also given.
         b) also shows the prediction of the GRS-LO parametrisation 
         for the structure function at $\qsq=20$~\gevsq
         and its point-like component separately.
        }\label{fig:fig04}
\end{center}
\end{figure}
%
\end{document}